\documentclass[aps,preprint,showpacs]{revtex4}
\usepackage{graphicx}

\newcommand{\E}{\mathbf{E}}
\newcommand{\inp}{\text{in}}
\newcommand{\out}{\text{out}}
\begin{document}
\title{Optical spin-to-orbital angular momentum conversion in inhomogeneous anisotropic media}
\author{L. Marrucci}
\email{lorenzo.marrucci@na.infn.it}
\author{C. Manzo}
\author{D. Paparo}
\affiliation{Dipartimento di Scienze Fisiche, Universit\`{a} di
Napoli ``Federico II'' and CNR-INFM Coherentia, Compl.\ di Monte
S.Angelo, v.\ Cintia, 80126 Napoli, Italy}
\date{March 3, 2006}
\begin{abstract}
We demonstrate experimentally an optical process in which the spin
angular momentum carried by a circularly polarized light beam is
converted into orbital angular momentum, leading to the generation
of helical modes with a wavefront helicity controlled by the input
polarization. This phenomenon requires the interaction of light with
matter that is both optically inhomogeneous and anisotropic. The
underlying physics is also associated with the so-called
Pancharatnam-Berry geometrical phases involved in any inhomogeneous
transformation of the optical polarization.
\end{abstract}
 \maketitle

\begin{figure}[h]
\includegraphics[angle=0, width=0.8\textwidth]{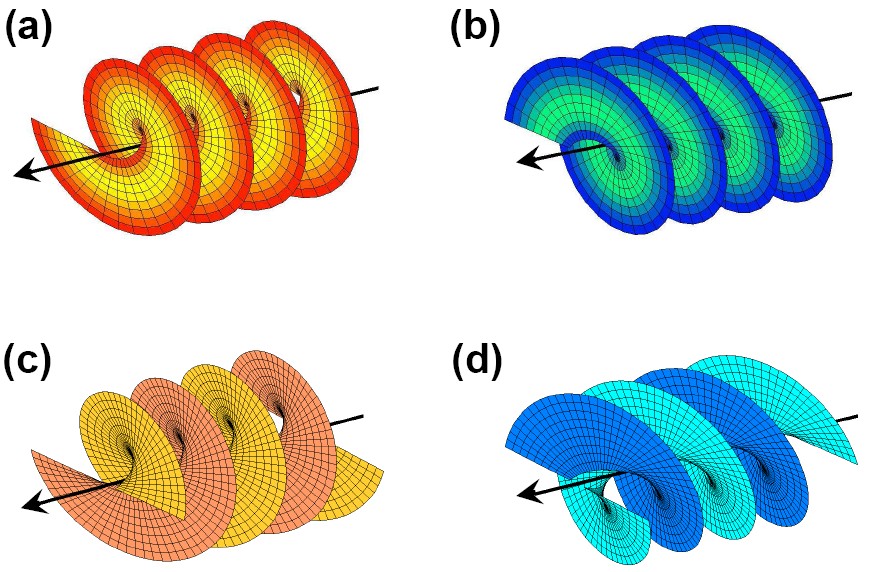}
\caption{Examples of helical waves. Represented are the wavefronts
of helical modes for helicities $m=+1$ (a), $m=-1$ (b), $m=+2$ (c),
and $m=-2$ (d).} \label{helixfig}
\end{figure}

A monochromatic light beam travelling along a given axis $z$ can
transport angular momentum oriented as the propagation direction in
two different forms \cite{humblet43,abbate91,allen92}. The first is
the classical ``spin-like'' form associated with the circular
polarizations of light, each photon carrying ${\pm}\hbar$ of angular
momentum depending on the handedness of the polarization. The second
is an ``orbital'' form, associated with the optical phase profile of
the beam in a plane orthogonal to the propagation axis, i.e.
parallel to the $xy$ plane. Using a complex notation
\cite{footnote1}, the electric field of a beam carrying a well
defined value of the orbital angular momentum can be written as
$\E(r,\varphi)=\E_0(r)\exp(im\varphi)$, where $r,\varphi$ are polar
coordinates in the $xy$ plane and $m$ is an integer. For such a
field, commonly named a ``helical mode'', it has been shown that
each photon carries a quantized intrinsic orbital angular momentum
($z$-component) given by $m\hbar$ \cite{allen92,allen99,oneil02}.
The wavefront of this field is composed of $|m|$ intertwined helical
surfaces, with a handedness given by the sign of $m$, as shown in
Fig.~\ref{helixfig}. Moreover, these fields present a topological
phase singularity (an ``optical vortex'') at the beam axis
\cite{nye74,bazhenov90}. For definiteness, in the following we will
refer to $m$ as the orbital-helicity of the beam (also called
``charge'' of the vortex). In general, any optical wave can be
decomposed in circularly polarized helical modes carrying well
defined values of both spin and orbital angular momentum. A
particularly important example is that given by the
Laguerre-Gaussian set of modes, which are exact eigenmodes of the
wave equation in the paraxial approximation (Helmholtz equation).

When light propagates in vacuum or in a homogeneous and isotropic
transparent medium, both spin and orbital angular momentum are
separately conserved. Both forms of angular momentums can be however
transferred to matter in suitable conditions. When a photon is
absorbed, for example, it transfers all its angular momentum to the
absorbing particle, both spin and orbital \cite{he95prl,simpson97}.
However, the two forms of optical angular momentum may couple to
different material degrees of freedom, when the particle is located
off axis \cite{oneil02}. Transfer of spin-only angular momentum to
matter is achieved in optically anisotropic media, as in normal
birefringent wave plates \cite{beth36}, trapped microscopic
particles \cite{friese98}, and liquid crystals
\cite{santamato86,abbate91,marrucci92}. An independent coupling of
orbital-only angular momentum with matter is instead possible in
inhomogeneous isotropic transparent media
\cite{beijersbergen93,beijersbergen94} (see also the related
demonstration of optical manipulation of particles with helical
beams \cite{paterson01}).

Clearly, a simultaneous independent coupling of both the spin and
orbital forms of angular momentum of light with matter is to be
expected in a medium which is both inhomogeneous and anisotropic
\cite{piccirillo04}. As we show in the following, these two separate
channels of angular momentum exchange between light and matter do
indeed exist in these optical media, but unexpectedly they are not
independent at all: in appropriate conditions, the exchange with
matter of the spin optical angular momentum affects the direction
(sign) of the exchange of the orbital angular momentum. There are
even specific geometries in which the two exchanges remain always
exactly opposite to each other, so that the total angular momentum
transferred to matter (the optical torque) vanishes identically. In
these cases, a direct transformation of the optical angular momentum
from the spin form to the orbital form takes place, with matter
playing only an intermediate role. We note that this is a rather
counterintuitive process, in which the input polarization of light
controls the shape of the output wavefront.

\begin{figure}[h]
\includegraphics[angle=0, width=0.6\textwidth]{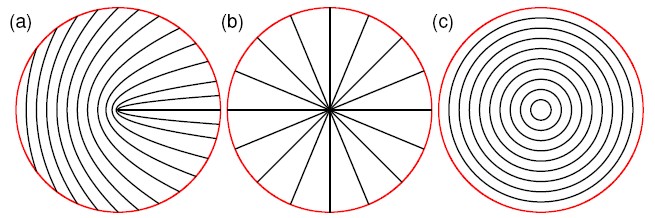}
\caption{Examples of $q$-plates. The tangent to the lines shown
indicate the local direction of the optical axis. (a) $q=1/2$ and
$\alpha_0=0$ (a nonzero $\alpha_0$ is here just equivalent to an
overall rigid rotation), which generates helical modes with
$m={\pm}1$; (b) $q=1$ with $\alpha_0=0$ and (c) with
$\alpha_0=\pi/2$, which can be both used to generate modes with
$m={\pm}2$. The last two cases correspond to rotationally symmetric
plates, giving rise to perfect spin-to-orbital angular momentum
conversion, with no angular momentum transfer to the plate.}
\label{qplatefig}
\end{figure}

To illustrate these effects, let us consider the specific case of a
planar slab of a uniaxial birefringent medium, having an homogeneous
birefringent phase retardation of $\pi$ (half-wave) across the slab,
and an inhomogeneous orientation of the fast (or slow) optical axis
lying parallel to the slab planes. We assume that the light beam
impinges on the slab at normal incidence, so that the slab planes
are parallel to the $xy$ coordinate plane. Moreover, the optical
axis orientation in the $xy$ plane, as specified by the angle
$\alpha$ it forms with the $x$ axis, is assumed to be given by the
following equation:
\begin{equation}\label{qplate}
\alpha(r,\varphi) = q\varphi + \alpha_0
\end{equation}
where $q$ and $\alpha_0$ are constants. Equation (\ref{qplate})
implies the presence of a defect in the medium localized at the
plane origin, $r=0$, similar to the typical defects spontaneously
formed by nematic liquid crystals \cite{degennes}. However, if $q$
is an integer or a semi-integer there will be no discontinuity line
in the slab. In the following, we will refer to inhomogeneous
birefringent elements having the above specified geometry as
$q$-plates. A few examples of $q$-plate geometries for different
values of $q$ and $\alpha_0$ are shown in Fig.~\ref{qplatefig}.

To analyze the effect of a $q$-plate on the optical field, it is
convenient to adopt a Jones formalism. The Jones matrix $\mathbf{M}$
to be applied on the field at each point of the $q$-plate transverse
plane $xy$ is the following:
\begin{equation}\label{qjones}
\mathbf{M}=\mathbf{R}(-\alpha)\cdot \left(
  \begin{array}{cc}
    1 & 0 \\
    0 & -1 \\
  \end{array}
\right) \cdot \mathbf{R}(\alpha) =
                \left(
                   \begin{array}{cc}
                     \cos2\alpha & \sin2\alpha \\
                     \sin2\alpha & -\cos2\alpha \\
                   \end{array}
                 \right),
\end{equation}
where $\mathbf{R}(\alpha)$ is the standard $2\times2$ rotation
matrix by angle $\alpha$ (in the $xy$ plane), and it is understood
that $\alpha$ depends on the point $(r,\varphi)$ in the transverse
plane according to Eq.~(\ref{qplate}).

A left-circular polarized plane wave, described by the Jones
electric-field vector $\E_{\inp}=E_0\times[1,i]$, after passing
through the $q$-plate, will be transformed into the following
outgoing wave (up to an unimportant overall phase shift):
\begin{equation}\label{outfield}
\E_{\out}=\mathbf{M}\cdot\E_{\inp} = E_0e^{i2\alpha}\left[
                                                            \begin{array}{c}
                                                              1 \\
                                                              -i \\
                                                            \end{array}
                                                          \right]=E_0e^{i2q\varphi}e^{i2\alpha_0}\left[
                                                            \begin{array}{c}
                                                              1 \\
                                                              -i \\
                                                            \end{array}
                                                          \right]
\end{equation}
The wave emerging from the $q$-plate is therefore uniformly
right-circular polarized, as would occur for a normal half-wave
plate, but it has also acquired a phase factor $\exp(im\varphi)$
with $m=2q$, i.e. it has been transformed into a helical wave with
orbital helicity $2q$ and orbital angular momentum $2q\hbar$ per
photon. It is easy to verify that in the case of a right-circular
input wave, the orbital helicity and angular momentum of the
outgoing wave are sign-inverted. In other words, the input
polarization of the light controls the sign of the orbital helicity
of the output wavefront. Its magnitude $|m|$ is instead fixed by the
birefringence axis geometry.

In passing through the plate, each photon being converted from
left-circular to right-circular changes its spin $z$-component
angular momentum from $+\hbar$ to $-\hbar$. In the case of a
$q$-plate having $q=1$, the orbital $z$-component angular momentum
of each photon changes instead from zero to $2\hbar$. Therefore, the
total variation of the angular momentum of light is nil and there is
no net transfer of angular momentum to the plate: the plate in this
case acts only as a ``coupler'' of the two forms of optical angular
momentum, allowing their conversion into each other. This exact
compensation of the spin and orbital angular momentum exchanges with
matter is clearly related to the circular symmetry (rotation
invariance) of the $q=1$ plate, as can be proved by general energy
arguments or by a variational approach to the optical angular
momentum fluxes \cite{abbate91}. If $q\neq1$, the plate is not
symmetric and will exchange an angular momentum of $\pm2\hbar(q -
1)$ with each photon, with a sign depending on the input
polarization. Therefore, in this general case the angular momentum
will not be just converted from spin to orbital, but the spin degree
of freedom will still control the ``direction'' of the angular
momentum exchange with the plate, besides the sign of the output
wavefront helicity.

To demonstrate these optical phenomena, we built a $q$-plate with
$q=1$ using nematic liquid crystal (LC) as the birefringent
material. The LC was sandwiched between two plane glasses, thus
forming a planar cell. The LC cell thickness (about 1 $\mu$m) and
material (E63 from Merck, Darmstadt, Germany) were chosen so as to
obtain a birefringence retardation of approximately a half wave, at
the working wavelength $\lambda=633$ nm. Before assembly, the inner
surfaces of the two glasses were coated with a polyimide for planar
alignment and one of them was briefly pressed against a piece of
fabric kept in continuous rotation. This procedure led to a surface
easy axis (i.e., the preferred orientation of LC molecules) having
the desired $q=1$ circular-symmetric geometry, as that shown in
Fig.~\ref{qplatefig}, panel (c).

\begin{figure}[h]
\includegraphics[angle=0, width=0.9\textwidth]{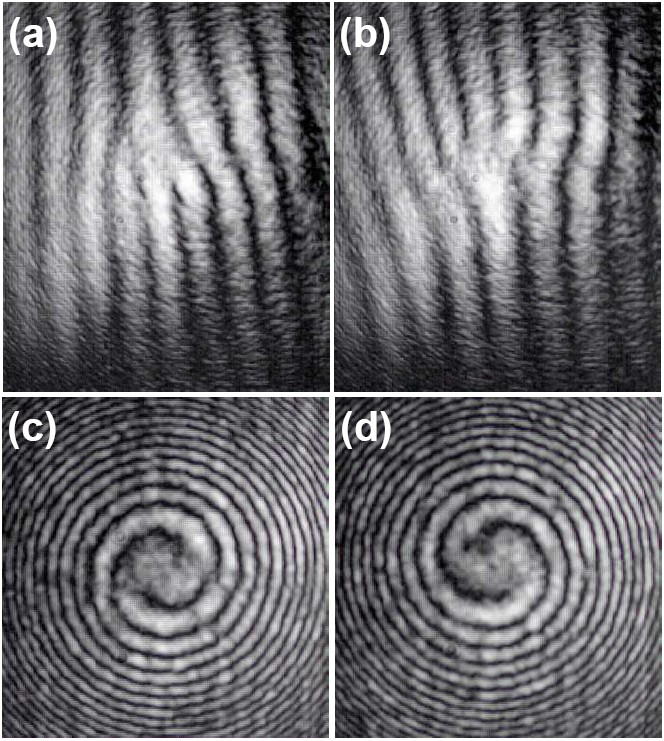}
\caption{Interference patterns of the helical modes emerging from
the LC cell (signal beam) after superposition with the reference
beam. Upper (a-b) panels refer to the plane-wave reference geometry,
lower (c-d) panels to the spherical-wave reference one. Panels on
the left, (a) and (c), are for a left-circular input polarization
and those on the right, (b) and (d), for a right-circular one.}
\label{spiralfig}
\end{figure}

In order to measure the wavefront shape of the light emerging from
the $q$-plate, we set up a Mach-Zender interferometer. A He-Ne laser
beam with a TEM$_{00}$ gaussian profile was split in two beams,
namely signal and reference. The signal beam was first circularly
polarized with the desired handedness by means of properly oriented
quarter-wave plate and then was sent through the LC $q$-plate. The
beam emerging from the $q$-plate was then sent through another
quarter-wave plate and a linear polarizer, arranged for transmitting
the polarization handedness opposite to the initial one, so as to
eliminate the residual unchanged circular polarization (this step is
not necessary when using a $q$-plate having exactly half-wave
retardation). Finally, the signal beam was superimposed with the
reference and thus generated an interference pattern directly on the
sensing area of a CCD camera. We used two different interference
geometries. In the first, the reference beam wavefront was kept
approximately plane (more precisely, it had the same wavefront
curvature as the signal beam) but the two beams were slightly tilted
with respect to each other. For non-helical waves, this geometry
gives rise to a regular pattern of parallel straight fringes. If the
wavefront of the signal beam is helical, the pattern develops a
dislocation (double, in this case, since $q=1$ yields $m=\pm2$),
with an orientation depending on the sign of $m$ and the relative
orientation of the two beams. In the second geometry, the reference
beam wavefront was approximately spherical, as obtained by inserting
a lens in the reference arm. For non-helical waves, the resulting
interference pattern is made of concentric circular fringes. If the
wavefront of the signal beam is helical, the pattern takes instead
the form of a spiral (a double spiral, for $m=\pm2$), with a
handedness depending on the sign of $m$ (counterclockwise outgoing
spirals, seen against the propagation direction as in our case,
correspond to a positive $m$). Figure \ref{spiralfig} shows the
CCD-acquired images of the interference patterns we obtained in the
two geometries, respectively for a left-circular (panels (a) and c))
and right-circular (panels (b) and (d)) input polarizations. These
results show unambiguously that the wavefront of the light emerging
from the $q$-plate is indeed helical with $m=\pm2$ (i.e. as shown in
panels (c) and (d) of Fig.~\ref{helixfig}), as predicted, and that
it carries an orbital angular momentum just opposite to the
variation of spin angular momentum associated with the polarization
occurring in the plate.

It must be emphasized that all commonly used methods for generating
helical modes of light (cilindrical lenses
\cite{allen92,beijersbergen93}, spiral plates \cite{beijersbergen94}
and holographic elements \cite{bazhenov90}) are associated
exclusively with an exchange of orbital angular momentum of light
with matter (indeed, they all involve inhomogeneous isotropic media)
and do not involve the wave polarization at all. In all these
methods the chirality of the medium structure is imprinted on the
generated wavefront, whose orbital helicity is therefore fixed
(although holographic spatial light modulators allow a slow
dynamical control of the generated helicity, by modifying the medium
spatial structure). In contrast, in the angular momentum conversion
process described here, the chirality of the generated wavefront is
determined by the input polarization handedness, and can therefore
be easily controlled dynamically.

The generation of electromagnetic helical waves based on spatially
non-uniform polarization transformations was previously reported by
Biener \textit{et al.}, who used subwavelength diffraction gratings
as birefringent elements and therefore were limited to the
mid-infrared spectral domain \cite{biener02,niv05}. However, Biener
\textit{et al.} did not discuss the conversion of optical angular
momentum involved in the process. Moreover, the interference
geometry adopted by Biener \textit{et al.} did not allow
distinguishing between right- and left-handed helical wavefronts and
therefore did not allow measuring the sign of the associated orbital
angular momentum. For this reason, the results reported here provide
to our knowledge the first actual demonstration of the all-optical
spin-to-orbital angular momentum conversion process, as well as the
first demonstration of the sign-inversion of the generated orbital
helicity occurring upon switching the input polarization handedness.

Biener \textit{et al.} highlighted another general physical
principle at the root of these optical phenomena: different
polarization transformations having the same initial and final
states involve optical phase differences of a geometrical nature,
known as Pancharatnam-Berry phases \cite{pancharatnam56,berry87}.
This principle is actually not limited to the generation of helical
waves, but it can be extended to any wavefront shaping controlled by
polarization transformations, thus setting the basis for an entirely
new approach to the design of phase optical elements
\cite{bhandari97,bomzon02}, an approach of which our work represents
the first actual demonstration in the visible domain.

In conclusion, we identified and experimentally demonstrated for the
first time an optical process in which the direct conversion of
optical angular momentum from the spin to the orbital form takes
place. This process leads to the generation of helical modes of
light with a wavefront helicity controlled by the input
polarization. This approach to the generation of helical modes of
light could prove particularly valuable in the foreseen applications
of these modes to the multi-state information encoding for classical
\cite{gibson04} and quantum communication and computation
\cite{molinaterriza02,leach02,vaziri02}, where the capability for a
fast switching of the generated helicity is critical.

We are grateful to Istvan J\'{a}nossy for a stimulating scientific
discussion which led us to conceiving this work, and to Enrico
Santamato and Giancarlo Abbate for lending us some experimental
equipment and for introducing us to the subject of the angular
momentum of light.


\begin{thebibliography}{29}
\expandafter\ifx\csname
natexlab\endcsname\relax\def\natexlab#1{#1}\fi
\expandafter\ifx\csname bibnamefont\endcsname\relax
  \def\bibnamefont#1{#1}\fi
\expandafter\ifx\csname bibfnamefont\endcsname\relax
  \def\bibfnamefont#1{#1}\fi
\expandafter\ifx\csname citenamefont\endcsname\relax
  \def\citenamefont#1{#1}\fi
\expandafter\ifx\csname url\endcsname\relax
  \def\url#1{\texttt{#1}}\fi
\expandafter\ifx\csname urlprefix\endcsname\relax\def\urlprefix{URL
}\fi \providecommand{\bibinfo}[2]{#2}
\providecommand{\eprint}[2][]{\url{#2}}

\bibitem[{\citenamefont{Humblet}(1943)}]{humblet43}
\bibinfo{author}{\bibfnamefont{J.}~\bibnamefont{Humblet}},
  \bibinfo{journal}{Physica} \textbf{\bibinfo{volume}{10}},
  \bibinfo{pages}{585} (\bibinfo{year}{1943}).

\bibitem[{\citenamefont{Abbate et~al.}(1991)\citenamefont{Abbate, Maddalena,
  Marrucci, Saetta, and Santamato}}]{abbate91}
\bibinfo{author}{\bibfnamefont{G.}~\bibnamefont{Abbate}},
  \bibinfo{author}{\bibfnamefont{P.}~\bibnamefont{Maddalena}},
  \bibinfo{author}{\bibfnamefont{L.}~\bibnamefont{Marrucci}},
  \bibinfo{author}{\bibfnamefont{L.}~\bibnamefont{Saetta}}, \bibnamefont{and}
  \bibinfo{author}{\bibfnamefont{E.}~\bibnamefont{Santamato}},
  \bibinfo{journal}{Phys.\ Scripta} \textbf{\bibinfo{volume}{T39}},
  \bibinfo{pages}{389} (\bibinfo{year}{1991}).

\bibitem[{\citenamefont{Allen et~al.}(1992)\citenamefont{Allen, Beijersbergen,
  Spreeuw, and Woerdman}}]{allen92}
\bibinfo{author}{\bibfnamefont{L.}~\bibnamefont{Allen}},
  \bibinfo{author}{\bibfnamefont{M.~W.} \bibnamefont{Beijersbergen}},
  \bibinfo{author}{\bibfnamefont{R.~J.~C.} \bibnamefont{Spreeuw}},
  \bibnamefont{and} \bibinfo{author}{\bibfnamefont{J.~P.}
  \bibnamefont{Woerdman}}, \bibinfo{journal}{Phys.\ Rev.\ A}
  \textbf{\bibinfo{volume}{45}}, \bibinfo{pages}{8185} (\bibinfo{year}{1992}).

\bibitem{footnote1}
We adopt the sign convention of the wave propagation factor
$\exp(ikz-i{\omega}t)$, which in turn fixes the sign-convention on
the wave orbital helicity and angular momentum.

\bibitem[{\citenamefont{Allen et~al.}(1999)\citenamefont{Allen, Padgett, and
  Babiker}}]{allen99}
\bibinfo{author}{\bibfnamefont{L.}~\bibnamefont{Allen}},
  \bibinfo{author}{\bibfnamefont{M.~J.} \bibnamefont{Padgett}},
  \bibnamefont{and} \bibinfo{author}{\bibfnamefont{M.}~\bibnamefont{Babiker}},
  \bibinfo{journal}{Prog.\ Opt.} \textbf{\bibinfo{volume}{39}},
  \bibinfo{pages}{291} (\bibinfo{year}{1999}).

\bibitem[{\citenamefont{O'Neil et~al.}(2002)\citenamefont{O'Neil, MacVicar,
  Allen, and Padgett}}]{oneil02}
\bibinfo{author}{\bibfnamefont{A.~T.} \bibnamefont{O'Neil}},
  \bibinfo{author}{\bibfnamefont{I.}~\bibnamefont{MacVicar}},
  \bibinfo{author}{\bibfnamefont{L.}~\bibnamefont{Allen}}, \bibnamefont{and}
  \bibinfo{author}{\bibfnamefont{M.~J.} \bibnamefont{Padgett}},
  \bibinfo{journal}{Phys.\ Rev.\ Lett.} \textbf{\bibinfo{volume}{88}},
  \bibinfo{pages}{053601} (\bibinfo{year}{2002}).

\bibitem[{\citenamefont{Nye and Berry}(1974)}]{nye74}
\bibinfo{author}{\bibfnamefont{J.~F.} \bibnamefont{Nye}} \bibnamefont{and}
  \bibinfo{author}{\bibfnamefont{M.~V.} \bibnamefont{Berry}},
  \bibinfo{journal}{Proc.\ Roy.\ Soc.\ Lond.\ A}
  \textbf{\bibinfo{volume}{336}}, \bibinfo{pages}{165} (\bibinfo{year}{1974}).

\bibitem[{\citenamefont{Bazhenov et~al.}(1990)\citenamefont{Bazhenov,
  Vasnetsov, and Soskin}}]{bazhenov90}
\bibinfo{author}{\bibfnamefont{V.~Y.} \bibnamefont{Bazhenov}},
  \bibinfo{author}{\bibfnamefont{M.~V.} \bibnamefont{Vasnetsov}},
  \bibnamefont{and} \bibinfo{author}{\bibfnamefont{M.~S.}
  \bibnamefont{Soskin}}, \bibinfo{journal}{JETP Lett.}
  \textbf{\bibinfo{volume}{52}}, \bibinfo{pages}{429} (\bibinfo{year}{1990}).

\bibitem[{\citenamefont{He et~al.}(1995)\citenamefont{He, Friese, Heckenberg,
  and Rubinsztein-Dunlop}}]{he95prl}
\bibinfo{author}{\bibfnamefont{H.}~\bibnamefont{He}},
  \bibinfo{author}{\bibfnamefont{M.~E.~J.} \bibnamefont{Friese}},
  \bibinfo{author}{\bibfnamefont{N.~R.} \bibnamefont{Heckenberg}},
  \bibnamefont{and}
  \bibinfo{author}{\bibfnamefont{H.}~\bibnamefont{Rubinsztein-Dunlop}},
  \bibinfo{journal}{Phys.\ Rev.\ Lett.} \textbf{\bibinfo{volume}{75}},
  \bibinfo{pages}{826} (\bibinfo{year}{1995}).

\bibitem[{\citenamefont{Simpson et~al.}(1997)\citenamefont{Simpson, Dholakia,
  Allen, and Padgett}}]{simpson97}
\bibinfo{author}{\bibfnamefont{N.~B.} \bibnamefont{Simpson}},
  \bibinfo{author}{\bibfnamefont{K.}~\bibnamefont{Dholakia}},
  \bibinfo{author}{\bibfnamefont{L.}~\bibnamefont{Allen}}, \bibnamefont{and}
  \bibinfo{author}{\bibfnamefont{M.}~\bibnamefont{Padgett}},
  \bibinfo{journal}{Opt.\ Lett.} \textbf{\bibinfo{volume}{22}},
  \bibinfo{pages}{52} (\bibinfo{year}{1997}).

\bibitem[{\citenamefont{Beth}(1936)}]{beth36}
\bibinfo{author}{\bibfnamefont{R.~A.} \bibnamefont{Beth}},
  \bibinfo{journal}{Phys.\ Rev.} \textbf{\bibinfo{volume}{50}},
  \bibinfo{pages}{115} (\bibinfo{year}{1936}).

\bibitem[{\citenamefont{Friese et~al.}(1998)\citenamefont{Friese, Nieminen,
  Heckenberg, and Rubinsztein-Dunlop}}]{friese98}
\bibinfo{author}{\bibfnamefont{M.~E.~J.} \bibnamefont{Friese}},
  \bibinfo{author}{\bibfnamefont{T.~A.} \bibnamefont{Nieminen}},
  \bibinfo{author}{\bibfnamefont{N.~R.} \bibnamefont{Heckenberg}},
  \bibnamefont{and}
  \bibinfo{author}{\bibfnamefont{H.}~\bibnamefont{Rubinsztein-Dunlop}},
  \bibinfo{journal}{Nature} \textbf{\bibinfo{volume}{394}},
  \bibinfo{pages}{348} (\bibinfo{year}{1998}).

\bibitem[{\citenamefont{Santamato et~al.}(1986)\citenamefont{Santamato, Daino,
  Romagnoli, Settembre, and Shen}}]{santamato86}
\bibinfo{author}{\bibfnamefont{E.}~\bibnamefont{Santamato}},
  \bibinfo{author}{\bibfnamefont{B.}~\bibnamefont{Daino}},
  \bibinfo{author}{\bibfnamefont{M.}~\bibnamefont{Romagnoli}},
  \bibinfo{author}{\bibfnamefont{M.}~\bibnamefont{Settembre}},
  \bibnamefont{and} \bibinfo{author}{\bibfnamefont{Y.~R.} \bibnamefont{Shen}},
  \bibinfo{journal}{Phys.\ Rev.\ Lett.} \textbf{\bibinfo{volume}{57}},
  \bibinfo{pages}{2423} (\bibinfo{year}{1986}).

\bibitem[{\citenamefont{Marrucci et~al.}(1992)\citenamefont{Marrucci, Abbate,
  Ferraiuolo, Maddalena, and Santamato}}]{marrucci92}
\bibinfo{author}{\bibfnamefont{L.}~\bibnamefont{Marrucci}},
  \bibinfo{author}{\bibfnamefont{G.}~\bibnamefont{Abbate}},
  \bibinfo{author}{\bibfnamefont{S.}~\bibnamefont{Ferraiuolo}},
  \bibinfo{author}{\bibfnamefont{P.}~\bibnamefont{Maddalena}},
  \bibnamefont{and}
  \bibinfo{author}{\bibfnamefont{E.}~\bibnamefont{Santamato}},
  \bibinfo{journal}{Phys.\ Rev.\ A} \textbf{\bibinfo{volume}{46}},
  \bibinfo{pages}{4859} (\bibinfo{year}{1992}).

\bibitem[{\citenamefont{Beijersbergen et~al.}(1993)\citenamefont{Beijersbergen,
  Allen, van~der Veen, and Woerdman}}]{beijersbergen93}
\bibinfo{author}{\bibfnamefont{M.~W.} \bibnamefont{Beijersbergen}},
  \bibinfo{author}{\bibfnamefont{L.}~\bibnamefont{Allen}},
  \bibinfo{author}{\bibfnamefont{H.~E. L.~O.} \bibnamefont{van~der Veen}},
  \bibnamefont{and} \bibinfo{author}{\bibfnamefont{J.~P.}
  \bibnamefont{Woerdman}}, \bibinfo{journal}{Opt.\ Commun.}
  \textbf{\bibinfo{volume}{96}}, \bibinfo{pages}{123} (\bibinfo{year}{1993}).

\bibitem[{\citenamefont{Beijersbergen et~al.}(1994)\citenamefont{Beijersbergen,
  Coerwinkel, Kristensen, and Woerdman}}]{beijersbergen94}
\bibinfo{author}{\bibfnamefont{M.~W.} \bibnamefont{Beijersbergen}},
  \bibinfo{author}{\bibfnamefont{R.~P.~C.} \bibnamefont{Coerwinkel}},
  \bibinfo{author}{\bibfnamefont{M.}~\bibnamefont{Kristensen}},
  \bibnamefont{and} \bibinfo{author}{\bibfnamefont{J.~P.}
  \bibnamefont{Woerdman}}, \bibinfo{journal}{Opt.\ Commun.}
  \textbf{\bibinfo{volume}{112}}, \bibinfo{pages}{321} (\bibinfo{year}{1994}).

\bibitem[{\citenamefont{Paterson et~al.}(2001)\citenamefont{Paterson,
  MacDonald, Arlt, Sibbett, Bryant, and Dholakia}}]{paterson01}
\bibinfo{author}{\bibfnamefont{L.}~\bibnamefont{Paterson}},
  \bibinfo{author}{\bibfnamefont{M.~P.} \bibnamefont{MacDonald}},
  \bibinfo{author}{\bibfnamefont{J.}~\bibnamefont{Arlt}},
  \bibinfo{author}{\bibfnamefont{W.}~\bibnamefont{Sibbett}},
  \bibinfo{author}{\bibfnamefont{P.~E.} \bibnamefont{Bryant}},
  \bibnamefont{and} \bibinfo{author}{\bibfnamefont{K.}~\bibnamefont{Dholakia}},
  \bibinfo{journal}{Science} \textbf{\bibinfo{volume}{292}},
  \bibinfo{pages}{912} (\bibinfo{year}{2001}).

\bibitem[{\citenamefont{Piccirillo and Santamato}(2004)}]{piccirillo04}
\bibinfo{author}{\bibfnamefont{B.}~\bibnamefont{Piccirillo}} \bibnamefont{and}
  \bibinfo{author}{\bibfnamefont{E.}~\bibnamefont{Santamato}},
  \bibinfo{journal}{Phys.\ Rev.\ E} \textbf{\bibinfo{volume}{69}},
  \bibinfo{pages}{056613} (\bibinfo{year}{2004}).

\bibitem[{\citenamefont{de~Gennes}(1974)}]{degennes}
\bibinfo{author}{\bibfnamefont{P.~G.} \bibnamefont{de~Gennes}},
  \emph{\bibinfo{title}{The Physics of Liquid Crystals}}
  (\bibinfo{publisher}{Oxford University Press}, \bibinfo{address}{Oxford},
  \bibinfo{year}{1974}).

\bibitem[{\citenamefont{Biener et~al.}(2002)\citenamefont{Biener, Niv, Kleiner,
  and Hasman}}]{biener02}
\bibinfo{author}{\bibfnamefont{G.}~\bibnamefont{Biener}},
  \bibinfo{author}{\bibfnamefont{A.}~\bibnamefont{Niv}},
  \bibinfo{author}{\bibfnamefont{V.}~\bibnamefont{Kleiner}}, \bibnamefont{and}
  \bibinfo{author}{\bibfnamefont{E.}~\bibnamefont{Hasman}},
  \bibinfo{journal}{Opt.\ Lett.} \textbf{\bibinfo{volume}{27}},
  \bibinfo{pages}{1875} (\bibinfo{year}{2002}).

\bibitem[{\citenamefont{Niv et~al.}(2005)\citenamefont{Niv, Biener, Kleiner,
  and Hasman}}]{niv05}
\bibinfo{author}{\bibfnamefont{A.}~\bibnamefont{Niv}},
  \bibinfo{author}{\bibfnamefont{G.}~\bibnamefont{Biener}},
  \bibinfo{author}{\bibfnamefont{V.}~\bibnamefont{Kleiner}}, \bibnamefont{and}
  \bibinfo{author}{\bibfnamefont{E.}~\bibnamefont{Hasman}},
  \bibinfo{journal}{Opt.\ Commun.} \textbf{\bibinfo{volume}{251}},
  \bibinfo{pages}{306} (\bibinfo{year}{2005}).

\bibitem[{\citenamefont{Pancharatnam}(1956)}]{pancharatnam56}
\bibinfo{author}{\bibfnamefont{S.}~\bibnamefont{Pancharatnam}},
  \bibinfo{journal}{Proc.\ Indian Acad.\ Sci.\ Sect.\ A}
  \textbf{\bibinfo{volume}{44}}, \bibinfo{pages}{247} (\bibinfo{year}{1956}).

\bibitem[{\citenamefont{Berry}(1987)}]{berry87}
\bibinfo{author}{\bibfnamefont{M.~V.} \bibnamefont{Berry}},
  \bibinfo{journal}{J. Mod.\ Opt.} \textbf{\bibinfo{volume}{34}},
  \bibinfo{pages}{1401} (\bibinfo{year}{1987}).

\bibitem[{\citenamefont{Bhandari}(1997)}]{bhandari97}
\bibinfo{author}{\bibfnamefont{R.}~\bibnamefont{Bhandari}},
  \bibinfo{journal}{Phys.\ Rep.} \textbf{\bibinfo{volume}{281}},
  \bibinfo{pages}{1} (\bibinfo{year}{1997}).

\bibitem[{\citenamefont{Bomzon et~al.}(2002)\citenamefont{Bomzon, Biener,
  Kleiner, and Hasman}}]{bomzon02}
\bibinfo{author}{\bibfnamefont{Z.}~\bibnamefont{Bomzon}},
  \bibinfo{author}{\bibfnamefont{G.}~\bibnamefont{Biener}},
  \bibinfo{author}{\bibfnamefont{V.}~\bibnamefont{Kleiner}}, \bibnamefont{and}
  \bibinfo{author}{\bibfnamefont{E.}~\bibnamefont{Hasman}},
  \bibinfo{journal}{Opt.\ Lett.} \textbf{\bibinfo{volume}{27}},
  \bibinfo{pages}{1141} (\bibinfo{year}{2002}).

\bibitem[{\citenamefont{Gibson et~al.}(2004)\citenamefont{Gibson, Courtial,
  Padgett, Vasnetsov, Pas'ko, Barnett, and Franke-Arnold}}]{gibson04}
\bibinfo{author}{\bibfnamefont{G.}~\bibnamefont{Gibson}},
  \bibinfo{author}{\bibfnamefont{J.}~\bibnamefont{Courtial}},
  \bibinfo{author}{\bibfnamefont{M.~J.} \bibnamefont{Padgett}},
  \bibinfo{author}{\bibfnamefont{M.}~\bibnamefont{Vasnetsov}},
  \bibinfo{author}{\bibfnamefont{V.}~\bibnamefont{Pas'ko}},
  \bibinfo{author}{\bibfnamefont{S.~M.} \bibnamefont{Barnett}},
  \bibnamefont{and}
  \bibinfo{author}{\bibfnamefont{S.}~\bibnamefont{Franke-Arnold}},
  \bibinfo{journal}{Opt.\ Express} \textbf{\bibinfo{volume}{12}},
  \bibinfo{pages}{5448} (\bibinfo{year}{2004}).

\bibitem[{\citenamefont{Molina-Terriza
  et~al.}(2002)\citenamefont{Molina-Terriza, Torres, and
  Torner}}]{molinaterriza02}
\bibinfo{author}{\bibfnamefont{G.}~\bibnamefont{Molina-Terriza}},
  \bibinfo{author}{\bibfnamefont{J.~P.} \bibnamefont{Torres}},
  \bibnamefont{and} \bibinfo{author}{\bibfnamefont{L.}~\bibnamefont{Torner}},
  \bibinfo{journal}{Phys.\ Rev.\ Lett.} \textbf{\bibinfo{volume}{88}},
  \bibinfo{pages}{013601} (\bibinfo{year}{2002}).

\bibitem[{\citenamefont{Leach et~al.}(2002)\citenamefont{Leach, Padgett,
  Barnett, Franke-Arnold, and Courtial}}]{leach02}
\bibinfo{author}{\bibfnamefont{J.}~\bibnamefont{Leach}},
  \bibinfo{author}{\bibfnamefont{M.~J.} \bibnamefont{Padgett}},
  \bibinfo{author}{\bibfnamefont{S.~M.} \bibnamefont{Barnett}},
  \bibinfo{author}{\bibfnamefont{S.}~\bibnamefont{Franke-Arnold}},
  \bibnamefont{and} \bibinfo{author}{\bibfnamefont{J.}~\bibnamefont{Courtial}},
  \bibinfo{journal}{Phys.\ Rev.\ Lett.} \textbf{\bibinfo{volume}{88}},
  \bibinfo{pages}{257901} (\bibinfo{year}{2002}).

\bibitem[{\citenamefont{Vaziri et~al.}(2002)\citenamefont{Vaziri, Weihs, and
  Zeilinger}}]{vaziri02}
\bibinfo{author}{\bibfnamefont{A.}~\bibnamefont{Vaziri}},
  \bibinfo{author}{\bibfnamefont{G.}~\bibnamefont{Weihs}}, \bibnamefont{and}
  \bibinfo{author}{\bibfnamefont{A.}~\bibnamefont{Zeilinger}},
  \bibinfo{journal}{Phys.\ Rev.\ Lett.} \textbf{\bibinfo{volume}{89}},
  \bibinfo{pages}{240401} (\bibinfo{year}{2002}).

\end{thebibliography}

\newpage

\newpage

\newpage

\end{document}